\documentclass[showpacs,aps,pre,superscriptaddress,twocolumn]{revtex4}
\usepackage{graphicx}
\usepackage{amsfonts}
\usepackage{color, verbatim}

\begin{document}
\title{Multidimensional Discrete Compactons in Nonlinear Schr\"odinger Lattices with Strong Nonlinearity Management}
\author{J. D'Ambroise}
\affiliation{ Department of Mathematics and Statistics, Amherst College, Amherst, MA, USA}
\author{M. Salerno }
\affiliation{Dipartimento di Fisica ``E.R. Caianiello'', CNISM and INFN
- Gruppo Collegato di Salerno, Universit\'a di Salerno, Via Giovanni
Paolo II, 84084 Fisciano (SA), Italy}
\author{P.G. Kevrekidis}
\affiliation{Department of Mathematics and Statistics, University of Massachusetts,
Amherst, MA USA}
\affiliation{Center for Nonlinear Studies and Theoretical Division, Los Alamos
National Laboratory, Los Alamos, NM 87544}
\author{F.Kh. Abdullaev}
\affiliation{ Department of Physics, Faculty of Sciences, IIUM, Jln. Indera Mahkota, Sultan Ahmad Shah, 25200,  Kuantan, Malaysia}
\affiliation{ CCNH, Universidade Federal do ABC, 09210-170, Santo Andr\'e, Brazil}

\begin{abstract}
The existence of multidimensional lattice compactons in the discrete nonlinear Schr\"odinger equation in the presence of fast periodic time modulations of the nonlinearity is demonstrated.  By averaging over the period of the fast modulations, a new effective averaged dynamical equation arises with coupling constants involving Bessel functions of the first and zeroth kind.  These terms allow one to solve, at this
averaged level, for exact discrete compacton solution configurations in the corresponding stationary equation.  We focus on seven types of compacton solutions: single site and vortex solutions are found to be always stable in the parametric regimes we examined.
Other solutions such as double site in- and out-of-phase, four site symmetric and anti-symmetric, and a five site compacton solution are found to have regions of stability and instability in two-dimensional parametric planes, involving
variations of the strength of the coupling and of the nonlinearity.  We also explore the time evolution of the solutions and compare the dynamics 
according to the averaged with those of the original dynamical equations
without the averaging.  Possible observation of compactons in the  BEC loaded in a deep two-dimensional optical lattice with interactions modulated periodically in time is discussed.
\end{abstract}

\maketitle
\newpage

\section{Introduction}
\label{intro}

Periodic management of parameters of  nonlinear lattices  
(and continua) is a very attractive technique for the generation of new types of systems or excitations with interesting localization  properties~\cite{Malomed_book}.  
For example, dynamic suppression of (atom and light, 
respectively) interwell tunneling 
has been reported experimentally in Bose-Einstein condensates~\cite{arimondo}
and in optical waveguide arrays~\cite{szameit}. 
On the other hand, very  fast periodic time  variation of the nonlinearity, also called  the {\it strong nonlinearity management} (SNLM) technique,  has been recently shown to be quite effective towards inducing compactly supported
(so-called compacton) solutions in lower dimensional discrete  systems such as one dimensional (1D) one and two component  Bose-Einstein condensates (BECs)
in optical lattices or arrays of nonlinear optical waveguides described  by the  discrete nonlinear Schr\"odinger (DNLS) equation \cite{AKS010,AHSU14}. The main feature  of such solutions, unlike  other types of  nonlinear excitations such as discrete breathers and intrinsic localized modes, is that their amplitudes decay to zero sharply without any tails. It has been shown for the 1D case that the lack of  exponential tails is a consequence of the nonlinear dispersive interaction induced by the SNLM which permits the vanishing of the intersite tunneling at compacton edges.  The absence of tails imply that compactons cannot interact with each other until they are in contact.  In the lattice
realm, this potentially allows for the possibility of ``maximal localization'' 
in the form of a single site solution.

Compactons are quite generic in nonlinear continua~\cite{rosenau,rosenau2} 
(in one and also higher~\cite{Ros1} dimensions) and
lattices~\cite{konotop1999,kevrekidis2002} (including compact
breathers~\cite{flach}) bearing {\it nonlinear dispersion}.  The difficulty of implementing this condition in physical  contexts has largely restricted 
the investigations mainly to the mathematical side, although this situation is 
now rapidly evolving.  Besides Bose-Einstein condensates under SNLM, 
physical   compactons were suggested to appear also in
exciton-polariton condensates \cite{konotop2013} and nearly-compact
(doubly exponential)
traveling waves have also been identified in the realm of granular
crystals bearing purely nonlinear interactions~\cite{atanas}.
Finally, another area of significant interest 
has recently arisen where special solutions in the
form of discrete compactons may emerge. This is due to the existence
of so-called flat bands in the linear dispersion relation
(due to the geometric characteristics of the corresponding lattice,
such as e.g. the Kagom{\'e} lattice)~\cite{rodrig1}. A realization of this 
type emerged very recently in the realm of the so-called Lieb
photonic lattices~\cite{Molina2015}.


Among the above different (Klein-Gordon, nonlinear Schr\"odinger,
Fermi-Pasta-Ulam, etc.) model equations in which compactly supported
structures have been proposed  DNLS 
ones are, arguably, of the most widely applicable
as models both for BECs and for nonlinear optics.
In the presence of SNLM, this type of system  supports discrete compactons 
via a site-dependent nonlinear rescaling of the inter-well tunneling (see \cite{AKS010} for details).  A similar approach,  applied to the quantum version of the DNLS model, i.e., to the Bose-Hubbard model with time dependent onsite interaction, has been shown to be quite effective for creating  new quantum phases in BECs with compacton-type excitations \cite{Rapp,Gres1} as well as for 
the generation of density dependent  synthetic gauge fields \cite{Gres2}.

These studies have mainly focused on the one dimensional case. On the other 
hand, it is known that compactons can  exist also in multidimensional contexts. In particular, multidimensional  compactons have been investigated in continuous models such as  (two-dimensional) variants of the
Korteweg--de Vries equations with nonlinear dispersion \cite{Ros1}, relativistic scalar field theories in two dimensions \cite{Bazeia}, 
Klein-Gordon and nonlinear Schr{\"o}dinger equations with sub-linear forces, etc..  In these last cases, however, the mechanism leading to compacton formation is not the nonlinear dispersion but the presence of a sub-quadratic interaction potential which  enforces compact patterns with sharp fronts \cite{Ros2,Ros3}. In this context, existence of vortex compactons was also demonstrated  to be possible, although only with a finite lifetime \cite{Ros3}. 

In the discrete case, multidimensional compactons have been scarcely investigated and only in the special case example of compact coherent
structures in the presence of a flat band of the linear 
spectrum~\cite{rodrig1,Molina2015}. 
To the best of our knowledge, general case examples 
(different numbers of sites, including ones bearing vorticity) of
multidimensional discrete compactons of the nonlinear Schr\"odinger type, 
amenable to  physical applications in BECs and nonlinear optics,  have 
not yet been investigated.


The aim of this paper is to demonstrate and systematically
explore the existence and stability of multidimensional  lattice 
compactons in the discrete nonlinear Schr\"odinger equation in the presence 
of  periodic time modulations of the nonlinearity i.e., in a 
2-dimensional generalization of~\cite{AKS010} bearing the potential
for a wide range of additional structures. In this work we concentrate 
mainly on compactons localized on no more than five interacting neighboring 
sites of a two-dimensional (2D) square lattice; of particular interest,
in addition to the stable single site compacton, is the stable four site 
compacton vortex solution i.e., bearing a vortical phase structure. 
In particular, the existence and stability properties of DNLS 
compacton excitations are investigated both for generic time dependent 
nonlinear modulations  and in  the SNLM limit for which an effective averaged DNLS model is derived. We  show that single site compactons are stable in the whole parameter space, however, two site compactons have finite stability 
ranges (although bearing some nontrivial differences from the ``standard''
DNLS model), different for symmetric and anti-symmetric types and with different dynamical features. Interestingly,  stationary three site compactons  with 
real amplitude cannot exist in the square geometry, while four site 
compactons of the symmetric or anti-symmetric types exist but have finite 
regions of instability similar to the double site configurations.  On  
four nearest-neighboring sites, however, for all values in the parameter space 
one has a stable vortex compacton in which the phase of the wavefunction increases by $\pi/2$  moving clockwise from one corner to the next of the square.  Since compactons interact only when they are in contact, one can obviously construct arbitrary single and two site compacton patterns by placing them on non interacting sites (such as for example next-neighboring sites along a diagonal). We also show that five site compactons with $C_4$ symmetry can exist and can be stable; their regions of instability are finite, similar to other 
configurations' instability.

The paper is organized as follows. In section II we introduce the model equation of a 2D DNLS with SNLM  and discuss the theoretical derivation of the effective averaged equations with nonlinear dispersion.  In section III we derive the existence conditions for exact compacton solutions of the averaged DNLS equation and in section IV we study numerically the linear stability properties and compare results with direct numerical integrations of the original (unaveraged) DNLS system. In the last section we discuss briefly the potential future experimental implementations, summarize our main results and provide some pointers for 
future work.

\section{Model and Theoretical Setup}
\label{thy}

Consider  the following 2D DNLS equation \cite{pgk_book}
\begin{eqnarray}
\label{dnls}
i \dot{u}_{n,m} &+&  k (u_{n+1,m} + u_{n-1,m}) { \ +  \ } \tilde k (u_{n, m+1} + u_{n,m-1}) \nonumber \\
&& + (\gamma_0 + \gamma(t))|u_{n,m}|^2 u_{n,m} = 0,
\end{eqnarray}
which serves as a model for the  dynamics of  BEC in optical lattices subjected to SNLM (through varying the interatomic
scattering length  by external time dependent  magnetic fields via a Feshbach resonance)~\cite{ABKS}, as well as for
light propagation in 2D optical waveguide arrays~\cite{Assanto}. In
the latter case, where this type of modulation has been realized not only
in discrete but also in continuum media (in both one and higher
dimensions)~\cite{pgk2}, the evolution variable is the propagation distance.
Hence, here the SNLM consists of periodic space variations of the Kerr 
nonlinearity along the propagation direction with the coupling constants 
$k, \tilde k$ quantifying the tunneling between adjacent sites along 
$n$ and $m$ directions, respectively, $\gamma_0$  denoting  the
onsite constant nonlinearity  and $\gamma(t)$ representing the time dependent 
modulation (of the interatomic interactions or of the 
refractive index, in atomic and optical settings, respectively). 
In the following we assume
a strong management case with $\gamma(t)$ being a periodic, e.g.  $\gamma(t)=\gamma(t+{T_0})$,  and rapidly varying function. As a prototypical example, 
we use $\gamma (t)=\frac {\gamma_1}\varepsilon \cos(\Omega \tau)$, 
with $\gamma_1 \sim O(1)$, $\varepsilon \ll 1$, $\tau=t/\varepsilon$ denoting the fast time variable and $T=2 \pi/\Omega$ 
the period {with respect to $\tau$ ($T_0=\varepsilon T$)}. In the following 
we take, for simplicity, the coupling constants $k, \tilde k$  to be the 
same for the two directions (assuming 
square symmetry): $k = \tilde k \equiv \kappa$.

The existence of compacton solutions of Eq. (\ref{dnls}) in the SNLM limit can be inferred  from (and analyzed in the context of) 
an effective  averaged 2D DNLS equation obtained by averaging out the fast time $\tau$. To that effect, 
it is convenient to  perform the transformation~\cite{JKP}
$u_{n,m}(t) = v_{n,m}(t)e^{i\Gamma |v_{n,m}(t)|^2}$ with $\Gamma  { \ =  \int_0^t dt\ \gamma (t) }= \gamma_1 \Omega^{-1} \sin(\Omega \tau)$,
which allows one to rewrite Eq. (\ref{dnls}) as
\begin{equation}
i \dot v_{n,m} = \Gamma v_{n,m} (|v_{n,m}|^2)_t - \kappa X {  \ - \ } \gamma_0 |v_{n,m}|^2 v_{n,m} ,
\label{ms1}
\end{equation}
with
\begin{eqnarray}
X&=& v_{n+1,m} e^{i \Gamma \theta^m_{+}} + v_{n-1,m} e^{i \Gamma \theta^m_{-}} + v_{n,m+1} e^{i \Gamma \theta^{+}_n} \nonumber \\ &&
 + v_{n,m-1} e^{i \Gamma \theta^{-}_n}, \mbox{ and }\\
\theta^m_{\pm} &=& |v_{n\pm 1,m}|^2- |v_{n,m}|^2, \;\;\theta^{\pm}_n= |v_{n,m \pm 1}|^2- |v_{n,m}|^2. \nonumber
\end{eqnarray}
On the other hand,  ${ ( |v_{n,m}|^2)_t=  (\dot v_{n,m} v_{n,m}^* + v_{n,m} \dot v_{n,m}^*)}= i \kappa (v_{n,m}^* X - v_{n,m} X^*)$, with the star denoting complex conjugation. Substituting this expression  into Eq. (\ref{ms1})  and averaging
the resulting equation over the period $T$ of the rapid modulation, we obtain
\begin{eqnarray}
i \dot v_{n,m} &=& i \kappa |v_{n,m}|^2 \langle \Gamma X \rangle - i \kappa v_{n,m}^2 \langle \Gamma X^* \rangle - \kappa \langle X \rangle \nonumber \\ &&  { \ - \ } \gamma_0 |v_{n,m}|^2 v_{n,m} ,
\label{ms2}
\end{eqnarray}
with $\langle \cdot \rangle \equiv \frac 1T \int_0^T ( \cdot ) d\tau$ denoting the fast time average. The averaged terms in Eq. (\ref{ms2}) can be calculated by means of the elementary integrals ${ \langle e^{ i \Gamma \theta_\pm} \rangle= J_0 (\alpha \theta_\pm)}$, $\; \langle \Gamma e^{\pm i \Gamma \theta_\pm} \rangle=\pm i \alpha J_1 (\alpha \theta_\pm)$,  with $J_i$ being Bessel functions of order $i=0,1$ and  $\alpha = \gamma_1/\Omega$, thus giving
\begin{eqnarray}
 i \dot{v}_{n,m} { = F(v)}
 \label{eqave}
 \end{eqnarray}
 with
 \begin{eqnarray}
&&  {  F(v)  } = - \gamma_0 |v_{n,m}|^2 v_{n,m}\nonumber \\
&&  - \alpha \kappa v_{n,m} [(v_{n+1,m}v_{n,m}^{\ast} + v_{n+1,m}^{\ast} v_{n,m}) J_1(\alpha\theta^{m}_{+}) \nonumber \\
&& + (v_{n,m+1}v_{n,m}^{\ast} + v_{n,m+1}^{\ast} v_{n,m}) J_1(\alpha\theta^{+}_n) \nonumber \\
&& + (v_{n-1,m}v_{n,m}^{\ast} + v_{n-1,m}^{\ast} v_{n,m})J_1(\alpha\theta_{-}^m)  \nonumber   \\
&& + (v_{n,m-1}v_{n,m}^{\ast} + v_{n,m-1}^{\ast} v_{n,m})J_1(\alpha\theta^{-}_n)]   \label{F}\\
&&- \kappa[v_{n+1,m} J_0(\alpha\theta_{+}^m) + v_{n,m+1} J_0(\alpha\theta^{+}_n)  \nonumber  \\
&& + v_{n-1,m} J_0(\alpha\theta_{-}^m)+v_{n,m-1} J_0(\alpha\theta^{-}_n)]. \nonumber
\end{eqnarray}
Note that the parameters $\gamma_1, \Omega  { \ \sim O(1)}$, and the averaged equation is valid for times $t \leq 1/\epsilon$.  This  modified DNLS equation can be written as $i \dot{v}_{n,m} = \delta H_{av}/\delta v_{n,m}^{\ast},$ with
the averaged Hamiltonian
\begin{eqnarray}
\label{ham}
&&
H_{av}= -\sum_{n,m} \{\kappa J_0(\alpha\theta_{+}^m) \left[v_{n+1,m}v_{n,m}^{\ast}+v_{n+1,m}^{\ast}v_{n,m} \right] \nonumber \\
&& + \kappa J_0(\alpha\theta^{+}_n) \left[v_{n,m+1}v_{n,m}^{\ast}+v_{n,m+1}^{\ast}v_{n,m} \right] +  \frac{\gamma_0}{2}|v_{n,m}|^4\}. \nonumber
\label{eqq7}
\end{eqnarray}
It is interesting to note that this Hamiltonian, except for the rescaling of the coupling constants  $k \rightarrow \kappa J_0(\alpha\theta_{+}^m)$, $\tilde k \rightarrow \kappa J_0(\alpha\theta^{+}_n)$,  is the same as the Hamiltonian of the DNLS equation in absence of SNLM (e.g. with $\gamma_1=0$).  A similar rescaling was reported  also for the 1D case  \cite{AKS010}.

It is also worth noting that while the appearance of the Bessel function is intimately connected with harmonic modulations, the existence of compacton solutions and the lattice tunneling suppression is generic for periodic SNLM.

The generalization of these equations to the case of 3D DNLS with cubic lattice symmetry is also quite straightforward to derive (omitted here for brevity).
In our numerical results below, we will restrict our considerations
to the numerically more tractable  
2D case (also more physically realistic
at least in the optics realm where the $z$-direction plays the
role of the propagation direction).

\section{Exact 2D compactons}
\label{excomp}

\begin{figure}
\begin{center}
\includegraphics[width=\columnwidth]{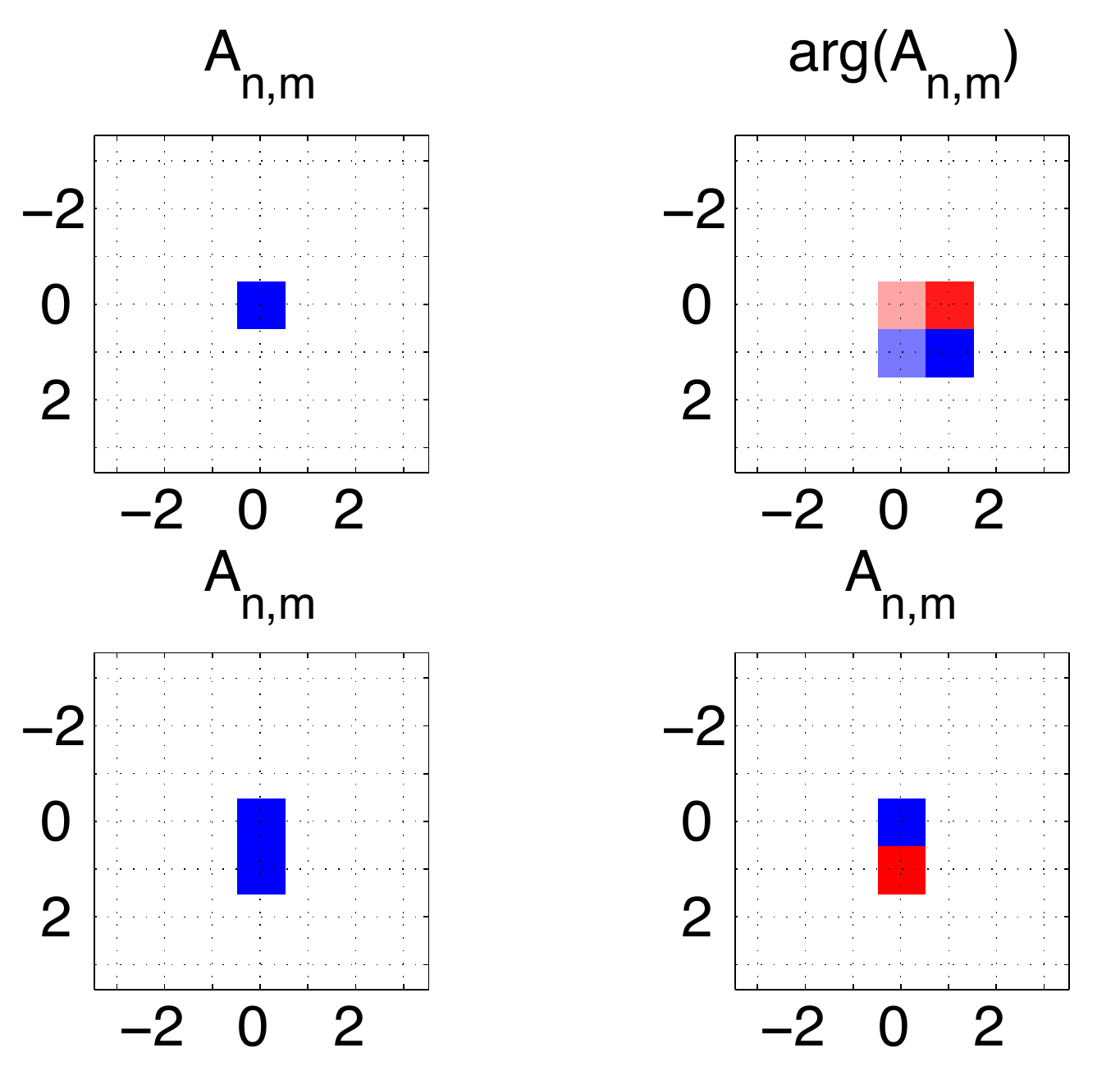}
\end{center}
\caption{(color online)
We plot typical examples of {\it exact} 
2D compacton solutions of Eq. (\ref{eqstat}).  In the top left panel we show a single site solution, in the bottom left we show a double-site in-phase solution, and in the bottom right, we show a double-site out-of-phase solution.  In these three plots colors show the (real) values of $A_{n,m}$ where blue represents $A_{n,m} = A_0$ and red $A_{n,m} = -A_0$.  No color represents zero amplitude.  In the top right plot we show a vortex solution where the four central colored sites all have amplitude $A_0$, and $arg(v_{n,m})\in\{-\pi/2,0,\pi/2,\pi\}$ is plotted by the four respective colors: dark red (at $(n,m)=(0,1)$), light red (at $(n,m)=(0,0)$), light blue (at $(n,m)=(1,0)$), dark blue (at $(n,m)=(1,1)$).  Here the length and width of the 2D grid are 
each $L = 7$.
}
\label{comp}
\end{figure}

To demonstrate the existence of {\it exact} compactons in the averaged system, we seek stationary solutions of the form
$v_{n,m} = A_{n,m} e^{-i \mu t}$ { with $A_{n,m}\in\mathbb{C}$} for which Eq. (\ref{eqave}) becomes
\begin{eqnarray}
 { \mu A_{n,m} = F(A)}
\label{eqstat}
\end{eqnarray}
 for $F$ given by Eq. (\ref{F}).  In the following, we  
theoretically predict and numerically verify that exact  solutions of this equation can exist in the form of genuine compactons i.e., 
possessing vanishing tails.

\begin{figure}
\begin{center}
\includegraphics[width=\columnwidth]{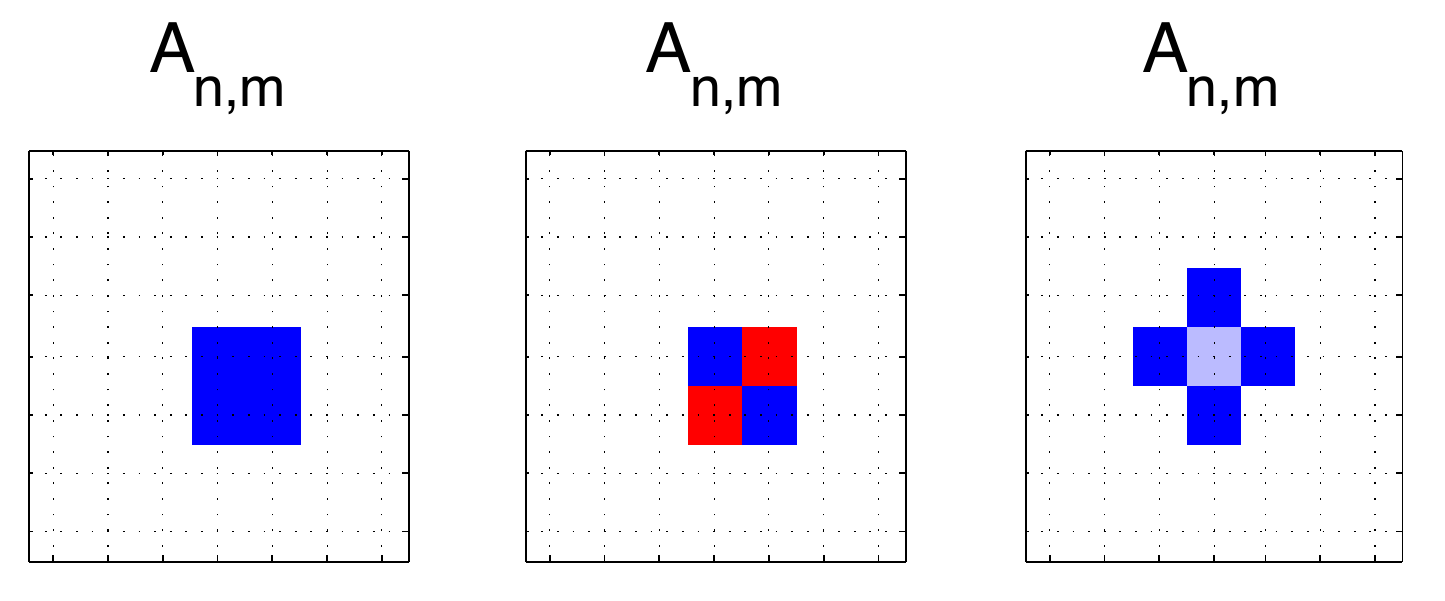}
\end{center}
\caption{(color online){
The plots show three additional examples of {\it exact} 
2D compacton solutions of Eq. (\ref{eqstat}) on four and five lattice sites.  In the left and middle panels symmetric and anti-symmetric four site compactons
are shown, respectively, while the right panel depicts a five site compacton.  Dark blue and red colors represent $A_{n,m} = A_0$ and  $A_{n,m} = -A_0$ similar to the single and double site plots in Figure \ref{comp}.  The light blue color at the center of the right-most plot represents the value $y$ for the center site of the five site solution obtained from Eq. (\ref{y5}).   Axis labels are the same here as those seen in Figure \ref{comp}.}}
\label{45}
\end{figure}

To search for compacton solutions, we begin by applying (\ref{eqstat}) at any zero-amplitude site that has exactly one out of its four neighbors nonzero.  We label this nonzero site $A_{N,M}\neq 0$ and we call it an ``edge site" of the compacton.  For $\kappa\neq 0$, this gives the condition
\begin{eqnarray}
J_0(\alpha |A_{N,M}|^2)=0 \quad \Rightarrow \quad |A_{N, M}|^2 = z_j/\alpha
\label{bessamp}
\end{eqnarray}
where $z_j$ is the $j^{th}$ zero of the Bessel function $J_0$.  In other words, each edge site of the compacton has amplitude
\begin{equation}
A_0 \stackrel{def.}{=} \sqrt{z_j/\alpha}.
\label{a0}
\end{equation}
Applying (\ref{eqstat}) at edge site(s) of the compacton gives additional condition(s) that depend on the shape of the compacton.  We will focus on the following shapes although other configurations are possible:  a single site, a double site in-phase, a double site out-of-phase, three types of four site solutions in the form of symmetric, asymmetric and vortex excitations, 
and finally a five site configuration.  Figures \ref{comp} and \ref{45} show one example of each of these configurations.  $L$ represents the horizontal (and vertical) length of the lattice.

\begin{figure}
\begin{center}
\includegraphics[width=\columnwidth]{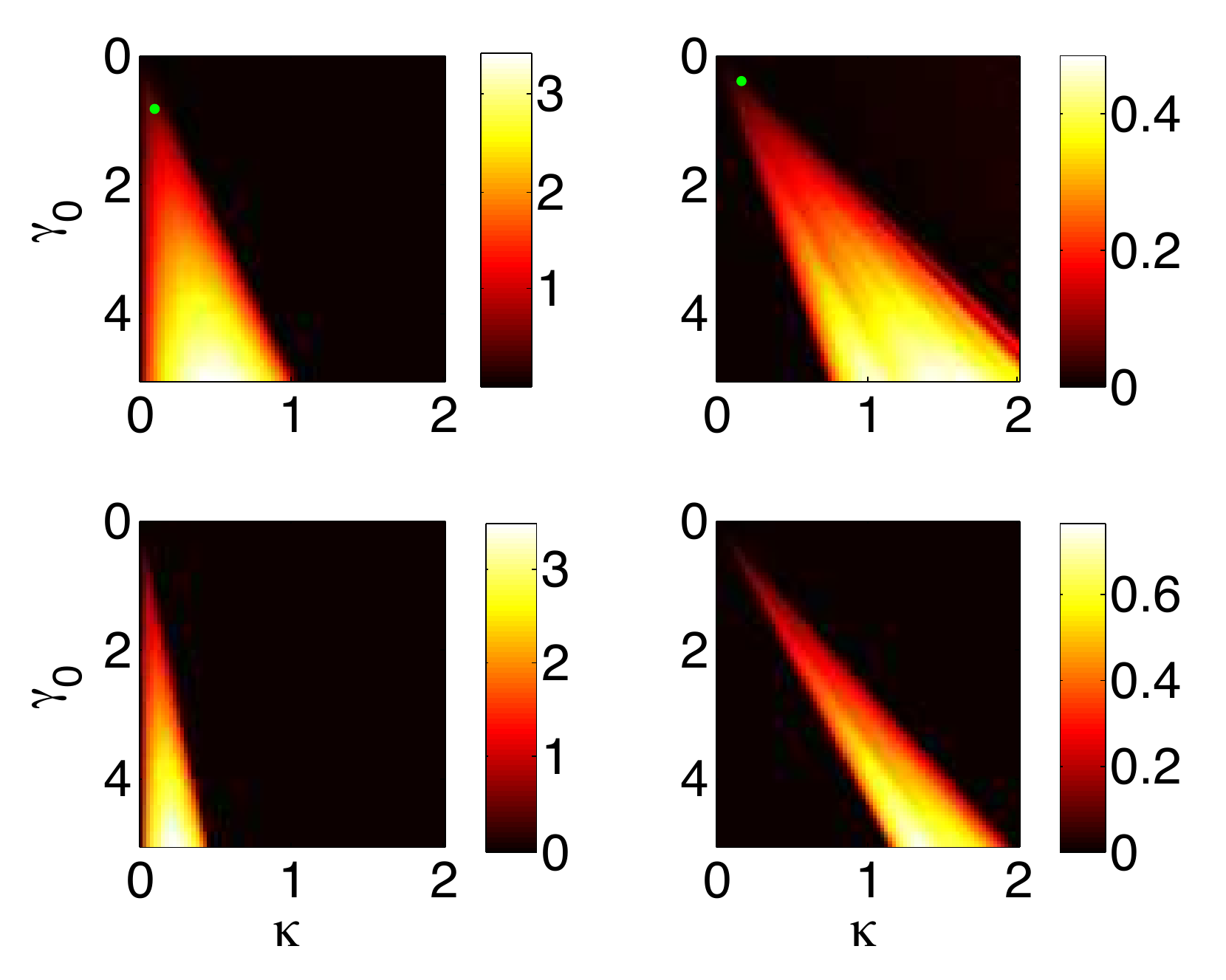}
\end{center}
\caption{(color online)
We plot $\max({\rm real}(\nu))$ in order to show the strength of the instabilities and regions of stability. Parameter values common across all four plots are $\epsilon = 0.01$ and $\Omega = \gamma_1 = 10$ so that $\alpha = 1$.  The left column of plots is associated with a double site in-phase solution with lattice length both horizontally and vertically $L=17$.  The right column of plots is associated with a double site out-of-phase solution with lattice length $L=15$.  The top row corresponds to $j=1$ and the bottom to $j=2$.  Notice that the top right plot shows more striations in the coloring than the top left plot; these further smooth out with higher lattice size $L$.  The green dots in the top two plots are parameter values for which the unstable solutions are propagated in time; see Figures \ref{dsi}, \ref{dso} below.
}
\label{stab}
\end{figure}

For a single site compacton solution located at $(n_0,m_0)$, this nonzero site is an edge site so it has amplitude $A_0$.  Taking $A_{n_0,m_0}=A_0$ and applying (\ref{eqstat})  at $(n_0,m_0)$ gives the value
\begin{eqnarray}
\mu=-\gamma_0 A_0^2.
\label{ssmu}
\end{eqnarray}
It is interesting to notice that this solution coincides with the one obtained for the one dimensional DNLS under SNLM in \cite{AKS010}.  This is due to the fact that for a single site the compact nature of the solution does not allow it to distinguish 1D from 2D or even 3D (this is true for a 3D cubic lattice as well).   Similar to the 1D case the single site compactons are stable; we will see this in more detail in Section \ref{stab}.

\begin{figure}
\begin{center}
\includegraphics[width=\columnwidth]{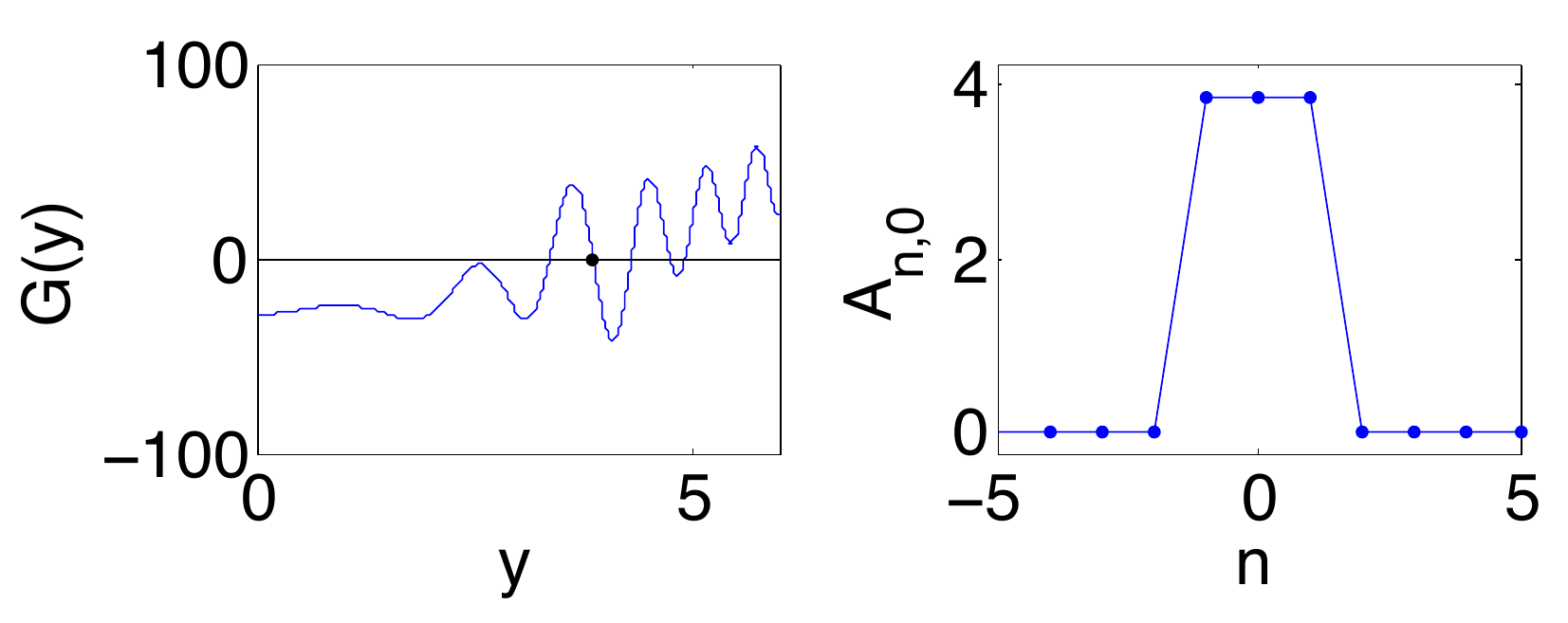}
\end{center}
\caption{(color online)
Graphical solution of Eq. (\ref{y5}) written as $G(y)=0$ (left panel) and section at $m=0$ of a five-site compacton (right panel)  of Eq. (\ref{eqstat}) for parameter values $\kappa=0.5, \alpha=1, \gamma_0=2$ with amplitudes at edges sites fixed in correspondence of the fifth zero, $z_5$, of the Bessel function $J_0$, e.g. $A_0=\sqrt{z_5/\alpha}\approx 3.864$. The amplitude of the central site is taken as  the zero of G(y) at  $y \approx 3.8695$ (black dot in the left panel) while the corresponding chemical potential is obtained from Eq. (\ref{mu5}) as $\mu \approx -30.678$.
}
\label{G5}
\end{figure}

For a double-site compacton located at $(n_0,m_0)$ and $(n_0 + 1,m_0)$ (or any other adjacent pair of indices) each with amplitude $A_0$ there are the two possibilties:  $A_{n_0+1,m_0}=\pm A_{n_0,m_0}$ where plus denotes an in-phase solution and minus an out-of-phase solution.  Using that $J_1(0)=0$ and $J_0(0)=1$, application of (\ref{eqstat}) at either of the two nonzero sites then gives the value
\begin{eqnarray}
\mu = \mp\kappa-\gamma_0A_{0}^2
\label{dsmu}
\end{eqnarray}
where the minus corresponds to the in-phase solution and the plus to the out-of-phase solution.

From the above cases one could expect that, although with different stability properties,  1D compactons located on the lattice along  $n$ or $m$ directions could be also solutions of multidimensional lattices.  This however,  except for the cases considered above, is not true in general. In this respect,  one and two-sites compactons are very special, since the number of edge sites in these solutions does not change with  dimensionality their analytical expressions remain  the same  in 1D, 2D and 3D. For  compactons with more than two sites, however, this is not true. Thus, for three-site compacton  with real amplitudes $(..., 0, A_1, A_2, A_1, 0,...)$, in the $n$ direction and zero amplitudes in all other sites, for example, one has  three  edge sites in 2D but only two in 1D.  In contrast with the 1D case, where solutions with  real amplitudes exist for arbitrary choices of parameters and are stable \cite{AKS010}, the constraints  resulting from Eq. (\ref{eqstat}) can be satisfied only for particular choices of parameters, e.g. real amplitude three-site compactons of 2D DNLS under SNLM are not  generic solutions in parameter space. The situation is different for complex 
and an example of this type will, in fact, be given below.

Genuine 2D compacton solutions first occur on four neighboring sites of a lattice cell. These square compactons can be either symmetric or anti-symmetric (see the panels of  Fig. \ref{45}).  In either case all four nonzero sites are edge sites with amplitude $A_0$.  Denoting with $(n_0, m_0)$ the coordinates of the upper left corner of the four site compacton, the symmetric solution has plus signs on the four nonzero sites so that 
\begin{eqnarray}
A_{n_0,m_0}=A_{n_0+1,m_0}=A_{n_0,m_0+1}=A_{n_0+1,m_0+1} = A_0. \nonumber
\end{eqnarray}
The anti-symmetric solution has an alternating pattern of plusses and minuses of the form
\begin{eqnarray}
A_{n_0,m_0}=A_{n_0+1,m_0+1} &=& A_0, \nonumber\\
 A_{n_0+1,m_0}=A_{n_0,m_0+1} &=& -A_0,  \nonumber
\end{eqnarray}
with  vanishing amplitudes, $A_{n,m}=0$, on all other sites. Substituting the above expressions
into Eq. (\ref{eqstat}) one readily gets the following equation for $\mu$ to be satisfied
\begin{eqnarray}
\pm 2 k + A_0^2 \gamma_0 + \mu =0
\end{eqnarray}
with the signs plus and minus referring to the symmetric and anti-symmetric case, respectively.

Quite remarkably, the DNLS system under SNLM  can also support 
discrete vortex-compactons, an unprecedented feature of 
earlier two-dimensional (continuous or discrete) compacton
considerations.  In particular, for a $2\times 2$ shaped vortex solution with the top left nonzero site located at index $(n_0,m_0)$, all four sites are edge sites with amplitude $A_0$.  Then we can write
\begin{eqnarray}
&A_{n_0,m_0} = A_0, \qquad  
& A_{n_0,m_0+1} = A_0e^{-i\pi/2}\nonumber\\ 
&A_{n_0+1,m_0}= A_0e^{i\pi/2}, \qquad 
&A_{n_0+1,m_0+1} = A_0e^{i\pi} 
\end{eqnarray}
and (\ref{eqstat}) applied on any of the four vortex sites gives
\begin{eqnarray}
\mu = - \gamma_0 A_0^2.
\end{eqnarray}

\begin{figure}
\begin{center}
\includegraphics[width=\columnwidth]{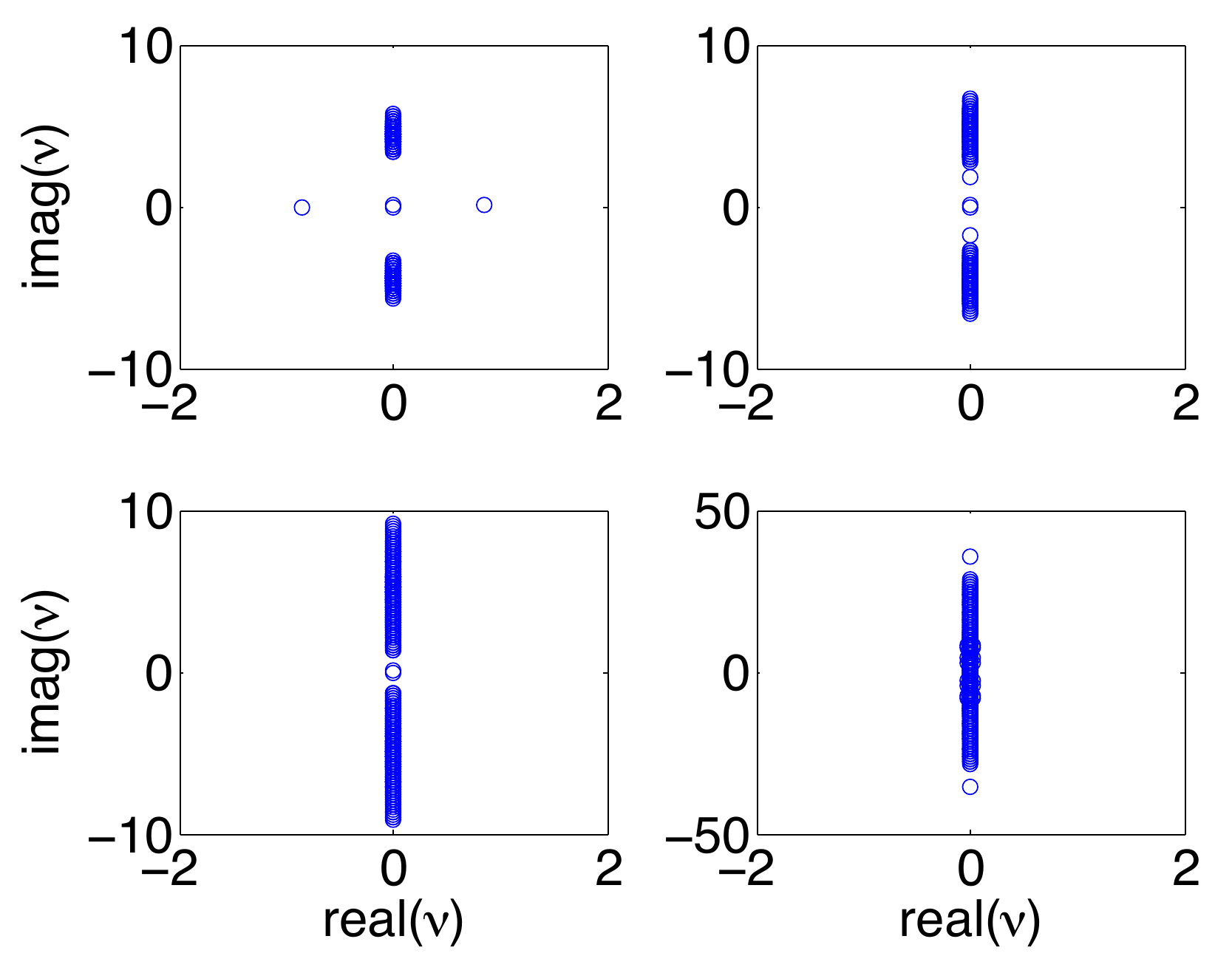}
\end{center}
\caption{(color online)
The panels show eigenvalues in the complex plane for selected parameters corresponding to a horizontal cut on the top left panel of Figure \ref{stab}
(in-phase solution), obtained by setting $\gamma_0 = 1.75$.    For $\kappa = 0$ (not shown) the eigenvalues are purely imaginary and four eigenvalues are zero.  For small nonzero $\kappa$ there are only two zero eigenvalues and the other two previously zero eigenvalues have begun to increase in magnitude on the real axis; this is shown for $\kappa = 0.3$ in the top left panel.  As $\kappa$ increases the solution stabilizes and these real eigenvalues move back towards zero horizontally along the real axis.  For even higher $\kappa$ values (at which the solution remains stable),
the two eigenvalues increase in magnitude again from zero, but this time along the imaginary axis; this is shown for $\kappa = 0.5$ in the upper right panel.  These eigenvalues merge with the band of eigenvalues on the imaginary axis meanwhile the (continuous spectrum) band lengthens; this is shown for $\kappa = 1$ in the bottom left panel.  At even higher $\kappa$ values, two complex conjugate purely imaginary eigenvalues emerge from the band at a magnitude higher than any eigenvalue in the band; this is depicted in the lower right panel for $\kappa = 5$.  The eigenvalues in the bottom right panel that appear to have very small nonzero real part are seen to actually have zero real part when the eigenvalues are computed for a much higher lattice length $L$.
}
\label{dsicut}
\end{figure}

\begin{figure}
\begin{center}
\includegraphics[width=\columnwidth]{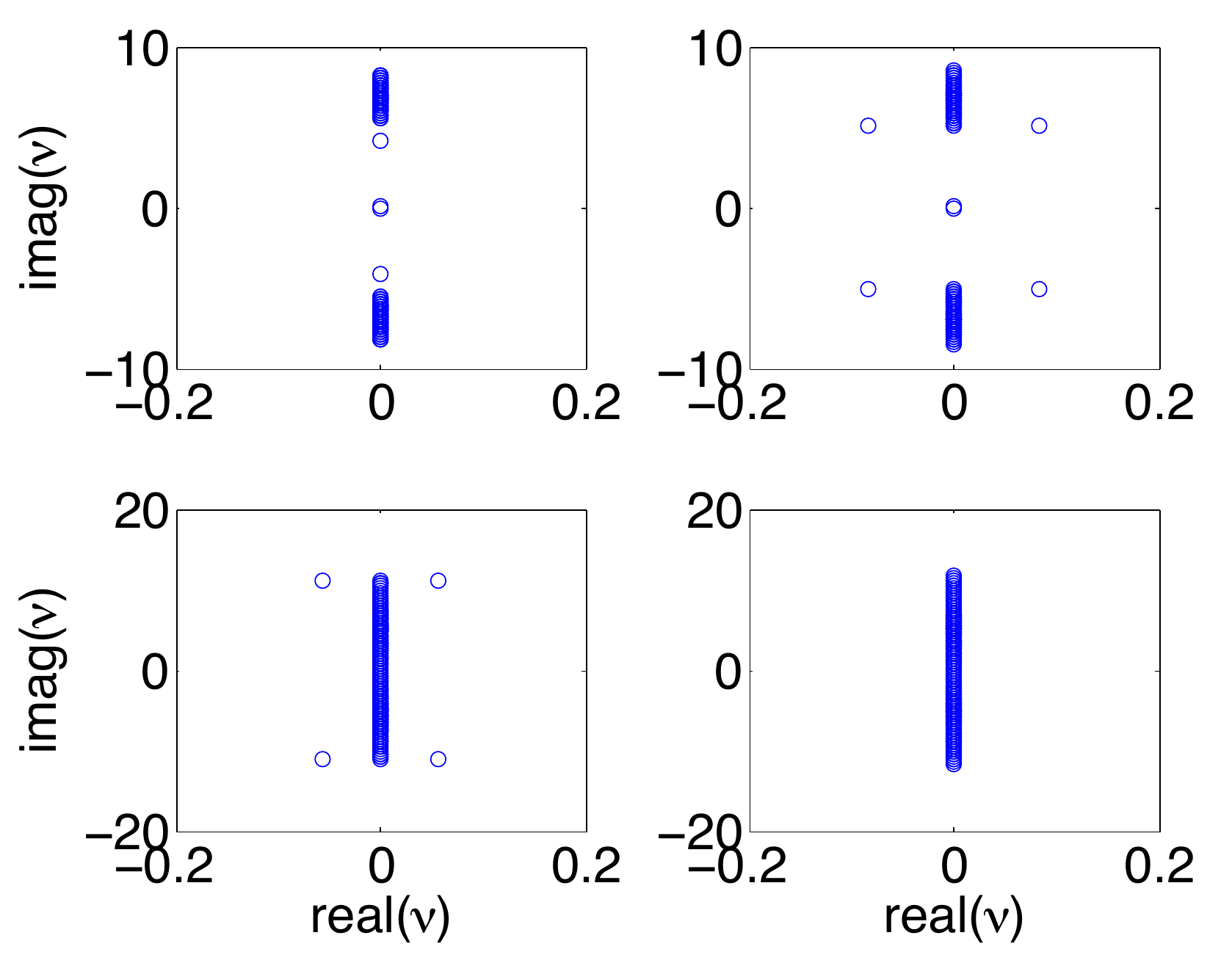}
\end{center}
\caption{(color online)
The panels show the eigenvalues in the complex plane for selected parameters corresponding to a horizontal cut on the top right panel of Figure \ref{stab}
(out-of-phase solution), obtained by setting $\gamma_0 = 3$.    For $\kappa = 0$ (not shown) the eigenvalues are purely imaginary and four eigenvalues are zero.  For small nonzero $\kappa$ there are only two zero eigenvalues and the other two previously zero eigenvalues have begun to increase in magnitude on the imaginary axis; this is shown for $\kappa = 0.35$ in the top left panel.  After the two merge with the (continuous spectrum) band of eigenvalues on the imaginary axis a quartet of complex eigenvalues then emerges for higher $\kappa$; this is depicted for $\kappa = 0.45$ in the top right panel.  For higher $\kappa$ values the four complex eigenvalues move in the complex plane, first increasing in real and imaginary parts then decreasing in real part until they merge again with the imaginary axis (meanwhile the continuous spectrum 
band on the imaginary axis lengthens); the bottom two plots show this restabilization with $\kappa = 1.38$ in the bottom left and $\kappa = 1.45$ in the bottom right.  At even higher $\kappa$ values (at which the solution remains stable) two complex conjugate purely imaginary eigenvalues emerge from the band at a magnitude higher than any eigenvalue in the band, similar to the lower right panel of Figure \ref{dsicut}.
}
\label{dsocut}
\end{figure}

With some additional  effort it is possible to obtain exact compacton solutions involving more ($>4$)  nonzero sites. As an example, consider the case of a five site compacton shaped as in the right  panel of Fig. \ref{45}.   Let $(n_0,m_0)$ denote the index of the center site of the configuration and define $y \equiv A_{n_0,m_0}$.  The four edge sites of the five site solution have amplitudes $A_0$ so we take $A_{n_0+1,m_0}=A_{n_0-1,m_0}=A_{n_0,m_0+1}=A_{n_0,m_0-1} = A_0$. Notice that the four edge sites (dark blue in the right panel of Fig. \ref{45}) isolate the central (light blue) site from the bulk, reproducing the same situation occurring for a three-site compacton in 1D (in 3D a similar solution would imply a  seven-site compacton).  Plugging the above edge site expressions into Eq. (\ref{eqstat}), applied at any one of the four edge sites, then gives an expression fixing the chemical potential
\begin{equation}
\mu(y) = - \kappa \frac{y}{A_0} J_0(\xi(y)) - \gamma_0 A_0^2 + 2\, \alpha\, \kappa \, A_0 \, y \,  J_1(\xi(y)),
\label{mu5}
\end{equation}
where $\xi(y)=\alpha(A_0^2 - y^2)$.  The final constraint  is obtained by applying Eq. (\ref{eqstat}) at the center site $(n_0,m_0)$.  Combining the result with the expression in Eq. (\ref{mu5}) for $\mu$ shows that the amplitude $y$ at the center site must be a solution of the equation $G(y)=0$ for
\begin{eqnarray}
 G(y)&\equiv& \kappa\left(\frac{4A_0}{y}-\frac{y}{A_0}\right) J_0(\xi(y)) - \frac{\gamma_0}{\alpha} \xi(y) \nonumber \\ 
&&   + 10\, \alpha \kappa\, A_0\, y\, J_1(\xi(y)).
 \label{y5}
\end{eqnarray}
Although  Eq. (\ref{y5}) does not yield a simple analytical expression for $y$,  it can be easily solved  numerically providing an exact five-site compacton with chemical potential given by Eq.~(\ref{mu5}). In general,  for fixed parameters and a fixed value for $A_0=\sqrt{z_n/\alpha}$, more than one zero can exist for G (see left panel of Fig. \ref{G5}), although not all of them 
necessarily lead to  stable compactons. It is remarkable, however, that some of them can have  non-vanishing intervals/ranges of stability in parameter space.  We now turn to the details of the stability considerations.

\section{Stability analysis and numerical simulations}
\label{stabsec}

\begin{figure}
\begin{center}
\includegraphics[width=\columnwidth]{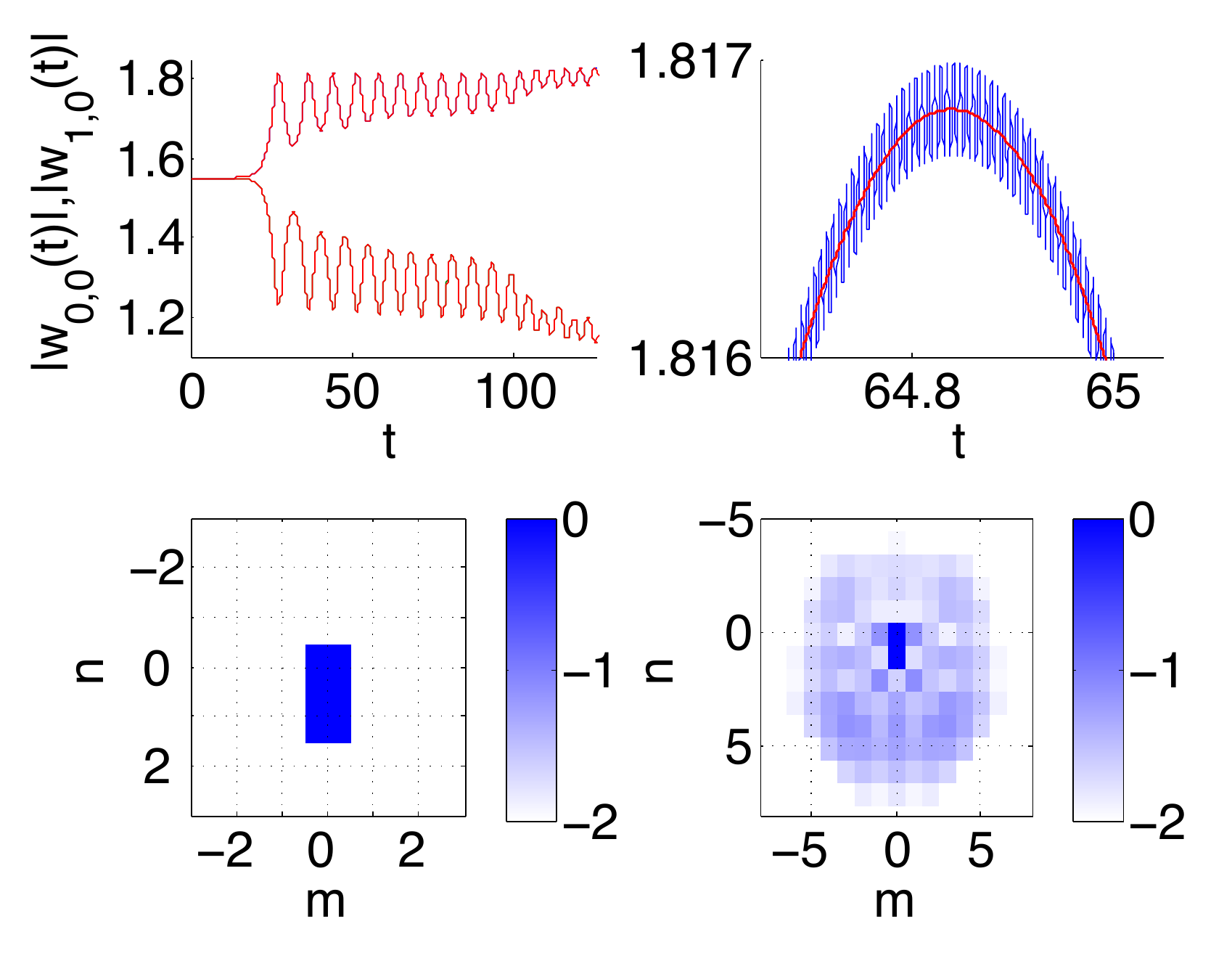}
\end{center}
\caption{(color online)
The plots show the result of propagation of the double site in-phase exact compacton solution in the time parameter $t$ according to each of the equations (\ref{dnls}) and (\ref{eqave}).  The values of $\epsilon, \Omega, \gamma_1, \alpha$ are the same as in Figure \ref{stab}; these values give $A_0\approx 1.5507$ for $j=1$ through Eq.~(\ref{a0}).  In the bottom left panel the plot shows $\log_{10}(|u_{n,m}(0)|)$ plotted as a function of $n,m$ where $u_{n,m}(0)=v_{n,m}(0)$ is the initial double site in-phase compacton configuration at $t=0$; this plot, of course, shows an order 0 logarithmic amplitude at the two 
excited sites and vanishing amplitude elsewhere.   The instability strength for this solution is $\max({\rm real}(\nu)) \approx 0.5327$.  In the bottom right panel the plot shows $\log_{10}(|v_{n,m}(120)|)$ where $v_{n,m}(t)$ is determined from $v_{n,m}(0)$ according to the averaged equation (\ref{eqave}); the plot of $\log_{10}(|u_{n,m}(120)|)$ determined from the non-autonomous equation (\ref{dnls}) is visually indistinguishable from the averaged version shown in the bottom right.  In the bottom right plot we see small amplitude excitations 
appearing near the compacton site, and these small amplitudes continue to spread out spatially as $t$ increases, while the central sites have amplitudes which approach nonzero values.  The top left panel shows this evolution of the amplitude of the two excited sites over time with the unaveraged magnitudes $|w_{0,0}(t)| = {|u_{0,0}(t)|}$ in blue and $|w_{1,0}(t)| = {|u_{1,0}(t)|}$ in green; these are propagated according to (\ref{dnls}).  The overlying red line plots correspond to the magnitudes $|w_{0,0}(t)| = {|v_{0,0}(t)|}$, $|w_{1,0}(t)| = {|v_{1,0}(t)|}$ propagated according to the averaged equation (\ref{eqave}).  From the perspective displayed in the upper left panel the averaged and the unaveraged amplitudes seem to lie on top of one another.   The top right panel here shows a small portion of the same plot zoomed in so as to show how the averaged solution in red in fact averages the non-autonomous equation solution (over a period
of the nonlinearity prefactor variation) in blue.
}
\label{dsi}
\end{figure}

\begin{figure}
\begin{center}
\includegraphics[width=\columnwidth]{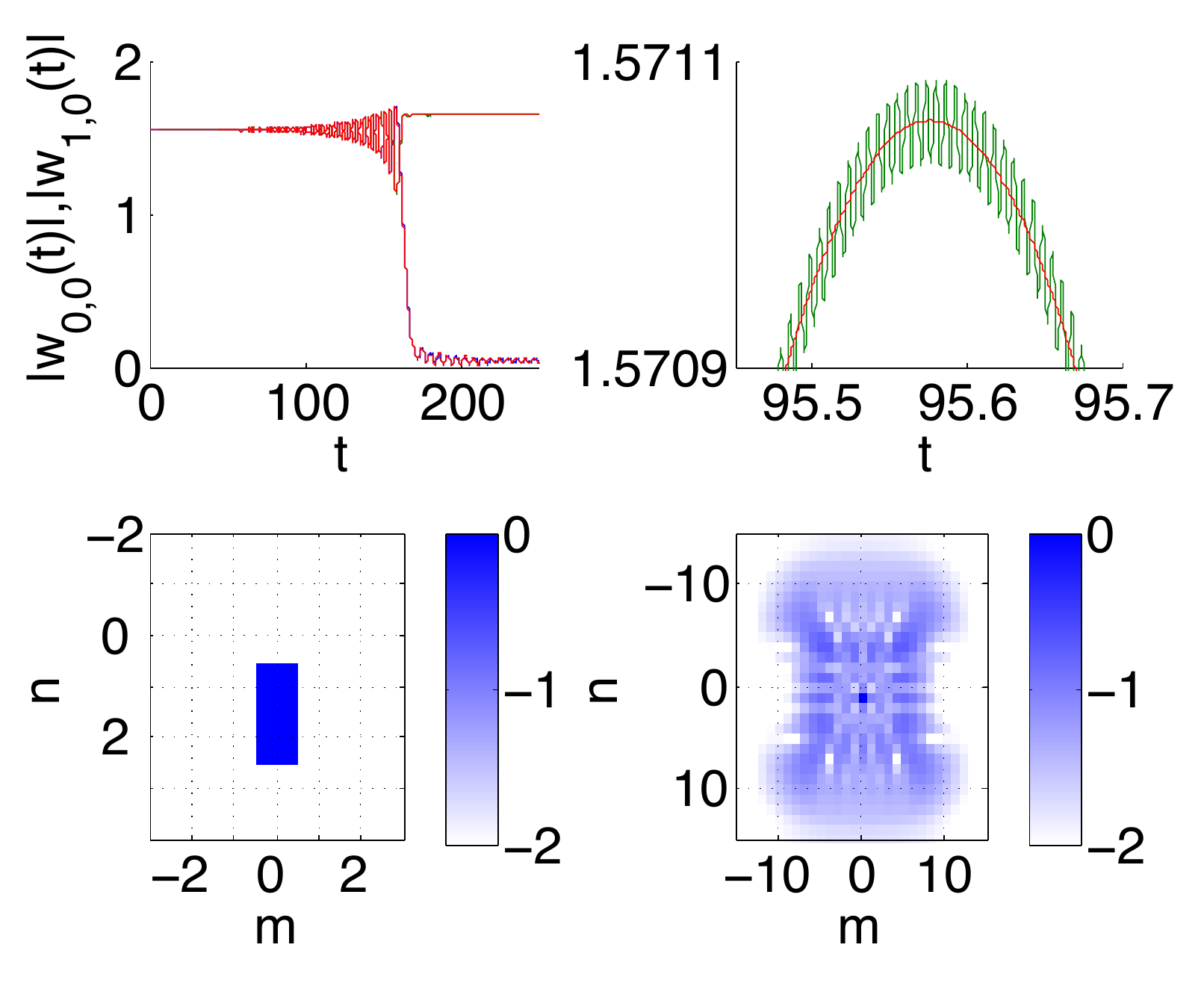}
\end{center}
\caption{(color online)
These plots are similar to those in Figure \ref{dsi} but here for the double site out-of-phase compacton solution; also the bottom right panel here corresponds to $t=180$.  The values of $ \epsilon, \Omega, \gamma_1, \alpha$ are the same as those seen in Figure \ref{stab} for the double site out-of-phase case.  These values again give $A_0\approx 1.5507$ for $j=1$ by (\ref{a0}).  The instability strength for this solution is $\max({\rm real}(\nu)) \approx 0.0298$. Notice that here, the eventual evolution leads to a single-site excitation (see the amplitude evolution in the top left and the eventual profile in the bottom right).
}
\label{dso}
\end{figure}

\begin{figure}
\begin{center}
\includegraphics[width=\columnwidth]{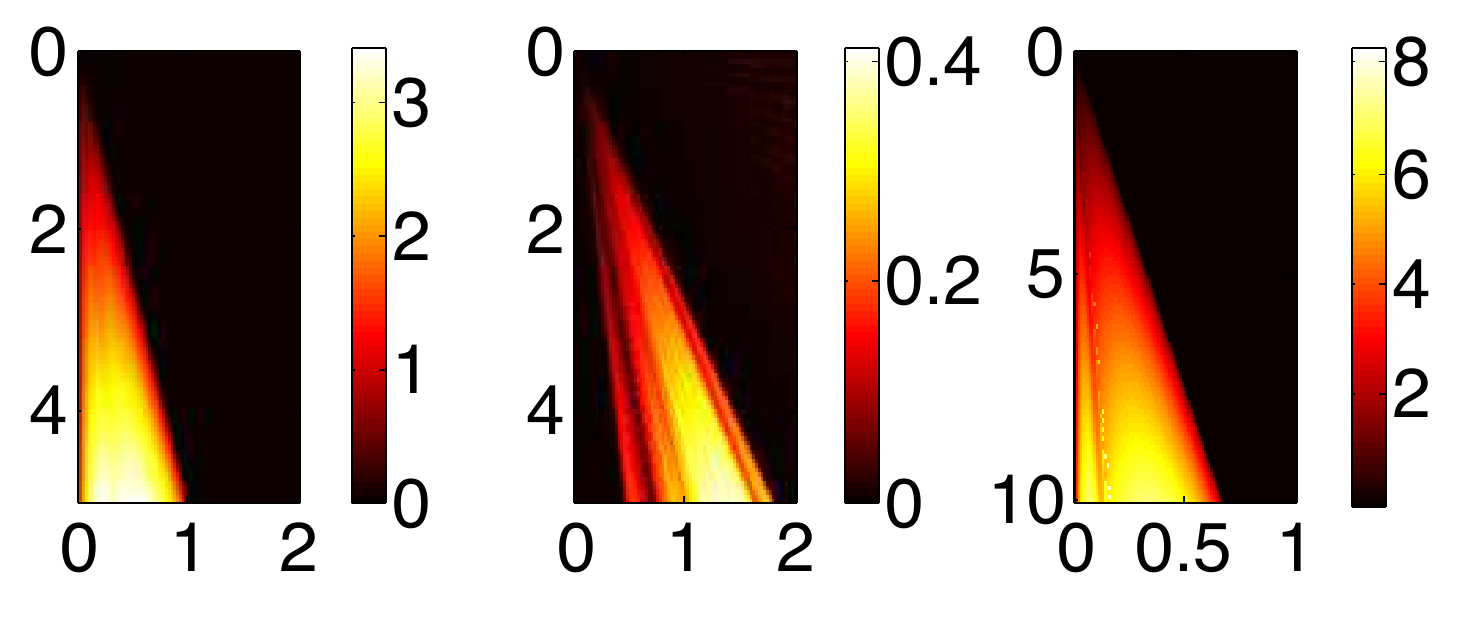}
\end{center}
\caption{(color online)
We plot $\max({\rm real}(\nu))$ in order to show the strength of the instabilities and regions of stability. Parameter values common across all four plots are $L=11$, $\epsilon = 0.01$ and $\Omega = \gamma_1 = 10$ so that $\alpha = 1$.  The left-most plot is associated to a four site symmetric solution using the first zero of the Bessel function with $j=1$, the middle plot corresponds to a four site asymmetric solution with  $j=1$, and the right-most plot is for the five site solution with $j=5$.  The axis labels are the same as those established in Figure \ref{stab}.
}
\label{45stab}
\end{figure}

In order to examine stability of the solutions found in Section \ref{excomp} we set
\begin{equation}
v_{n,m}(t) = \left( A_{n,m} + \delta e^{\nu t} \psi_{n,m} \right)e^{-i\mu t}
\end{equation}
where both $A_{n,m}$ and $\psi_{n,m}$ are independent of time $t$.  One can then show that for $\psi_{n,m} = a_{n,m}+ ib_{n,m}$ the real vector $\left( \begin{array}{c}a_{n,m}\\ b_{n,m}\end{array} \right)$ (with $2L^2$ components) is an eigenvector of the $2L^2\times 2L^2$ matrix
\begin{equation}
M = \left[ \begin{array}{cc}
-\frac{\partial Im(F(\psi))}{\partial a} & -\frac{\partial Im(F(\psi))}{\partial b} \\
&\\
\frac{\partial Re(F(\psi))}{\partial a} & \frac{\partial Re(F(\psi))}{\partial b}
\end{array}\right]
\end{equation}
with eigenvalue $\nu$.

We find that the single site and vortex solutions are stable for all parameter values $\kappa, \lambda_0$ so that $\max({\rm real}(\nu)) = 0$; in other words, the nonzero sites' time evolution plots according to the averaged Eq. (\ref{eqave}) even under perturbation does not lead to growth.  The time evolution of the single site or vortex solution according to the original Eq. (\ref{dnls}) 
with the time-periodic nonlinearity gives the expected result of oscillations about the stable solution given by the averaged equation.    The double site solutions have regions of instability which depend on the system's parameters.  
Figure \ref{stab} shows the maximal growth rate of the instability 
as a function of the dc strength of the nonlinearity and of the 
coupling constant. We see that the in-phase solutions present instability
intervals near the uncoupled limit (but contrary to what is the case
in the regular DNLS, these instability intervals end at a finite value
of the coupling). On the other hand, for the out-of-phase solutions,
the instability intervals arise for an intermediate range of the couplings,
a feature to which we will return below. Interestingly,
we also find that an increase in the index $j$ of the zero of the Bessel function $J_0$ decreases the size of the region of instability.  Additional analysis 
has shown that for the double site in-phase, an increase in the parameter 
$\alpha$ (i.e. decrease in frequency $\Omega/2\pi$ of the $\gamma$ function) 
decreases the size of the region of instability when plotted versus 
the coupling $\kappa$, and the dc nonlinearity strength
$\gamma_0$ as in Figure \ref{stab}.  For the double site out-of-phase, an increase in the parameter $\alpha$ (i.e. decrease in frequency $\Omega/2\pi$ of the $\gamma$ function) expands the region of instability  when plotted versus $\kappa,\gamma_0$ as in Figure \ref{stab}.

The nature of the double site instabilities is more
clearly demonstrated in Figures \ref{dsicut}, \ref{dsocut}.  For the double site in-phase solutions, unstable eigenvalues constitute a 
pair of equal in magnitude and opposite real numbers; see Figure \ref{dsicut} where more details of the transition between stable and unstable are discussed.  A key feature of the double site in-phase (in)stability is that for fixed $\gamma_0$, small nonzero $\kappa$ values give unstable solutions and higher $\kappa$ values stabilize the solution, a feature that is typically {\it absent}
in the standard DNLS model.   For the double site out-of-phase solutions, unstable eigenvalues appear as a quartet (due to the Hamiltonian nature of
the system), i.e. two complex conjugate pairs with equal magnitude plus/minus real parts.  This arises from the collision of the two imaginary eigenvalues
(in this case), stemming from the origin with the continuous spectrum,
a feature that is common in such out-of-phase focusing nonlinearity
settings; see e.g.~\cite{pgk_book}. 
More details of the transitions between stable and unstable solutions for the out-of-phase solutions are discussed in Figure \ref{dsocut}.  A key feature of the double site out-of-phase (in)stability is that for fixed $\gamma_0$, both small nonzero and large positive $\kappa$ values 
give stable solutions with the unstable region lying in a finite $\kappa$ 
interval that lies away from $\kappa = 0$. For large $\kappa$, the
formerly unstable quartet grows (in imaginary part) faster than the
band of the continuous spectrum and eventually returns to the imaginary
axis, leading to spectral stability.

For stable solutions in the averaged model the amplitudes of the excited sites
remain steady at the predicted value $A_0$ in (\ref{a0}) or
oscillate around it, if perturbed; this is the same feature described above for stable single site and vortex type solutions.  For stable solutions in the 
original non-autonomous model the amplitudes oscillate about the value 
$A_0$ as expected.  

In Figure \ref{stab} the green dots show select parameter values for 
which the dynamics of unstable double site solutions will be 
explored below.  Recall that in the case of the double-site solutions, 
equation (\ref{a0}) gives the amplitudes of the two excited sites $u_{0,0}, u_{1,0}$ at $t=0$.  In Figures \ref{dsi}, \ref{dso} plots of the evolution of the amplitudes of the unstable double site solutions over the propagation parameter $t$ are shown in detail, according to the averaged equation (\ref{eqave}) and the original non-autonomous equation (\ref{dnls}).   The unstable double site solutions (both in-phase and out-of-phase) transition to a non-compacton state 
with oscillating phase so that the initial phase profile is not preserved 
over time.  For the double site in-phase solution in Figure \ref{dsi} the two initially excited sites remain of order one amplitude as $t$ increases
(although the solution mass is asymmetrically distributed between
them due to their different amplitudes after the instability
manifestation). For sites in the vicinity of the original compact
support, there is gain in small amplitudes increasing the footprint of the solution.  For the double site out-of-phase solution in Figure \ref{dso} as $t$ increases the amplitude at the two center sites first oscillates and then one of the sites' amplitudes drops down towards zero; the resulting configuration is a non-compacton solution with order one magnitude only at one site, i.e., the
waveform degenerates towards a fundamental, single-site solution.

In Figure \ref{45stab} we show plots of the dependence 
of the instability strength (i.e., of the maximal growth
rate of the potential unstable modes) on the parameter grid for the four and five site solutions. The stability properties of both the four site symmetric 
(in-phase) compacton and the five site compacton configurations follow a similar pattern to that of the double site in-phase solutions; compare the left-most and right-most plots of Figure \ref{45stab} to the left column of Figure \ref{stab}.  That is, for fixed $\gamma_0$ as $\kappa$ increases from zero the solutions are immediately unstable with real eigenvalues and then they 
eventually stabilize for higher values of the coupling 
$\kappa$.  The eigenvalues for the unstable solutions appear in real pairs 
similar to that which is seen in Fig. \ref{dsicut}.  While the double site 
in-phase solutions have only one pair of real nonzero eigenvalues, the 
four site symmetric solutions have three pairs of real nonzero eigenvalues and 
the five site unstable solutions have four pairs. This is
in line with the expectations from the standard DNLS model~\cite{pgk_book}.

The stability properties of the four site anti-symmetric solutions 
(with adjacent sites being out-of-phase) follow a pattern similar to that of 
the double site out-of-phase solutions, where for small enough $\kappa$ the solutions are stable.  Increasing $\kappa$ one finds a finite (intermediate
coupling) interval of instability, and then for higher $\kappa$ the solution 
stabilizes again.  To see the similarity between the four site 
anti-symmetric solution stability and the double site out-of-phase stability compare the center plot of Figure \ref{45stab} to the right column of 
Figure \ref{stab}.  Plots of the eigenvalues in the complex plane look 
similar to the plots in Fig. \ref{dsocut}.  While the double site out-of-phase 
solutions feature for small coupling a single imaginary eigenvalue
pair and thus have only one potential quartet of unstable eigenvalues, 
the four site anti-symmetric solutions have three 
imaginary pairs for small $\kappa$ and may eventually feature up to
three eigenvalue quartets, again in line with what one may expect
in the standard DNLS case~\cite{pgk_book}.

\section{Conclusions \& Future Challenges}

Summarizing the findings of the present work, 
we have illustrated through a combination of analytical considerations
and numerical results the existence of 2D compactons of the discrete nonlinear Schr\"odinger equation in the presence of fast periodic time modulations of the nonlinearity. In particular, we showed that single site multidimensional compactons are very robust excitations,  two-site stationary compactons, both of symmetric and anti-symmetric type,  are also quite generic and can be stable in a wide region of the parameter space (of the coupling and nonlinearity prefactors). 
Four-site compactons have been found to be always stable only in the vortex state, an unusual feature which is fundamentally distinct from the case
of the standard discrete nonlinear Schr{\"o}dinger model. 
The five site compactons may also feature instabilities but can 
be controllably again stabilized for suitable parametric intervals
of the tunneling constant and/or dc-nonlinearity strength. 
These findings were not only obtained for the effective averaged DNLS equation 
that was derived herein, but were also confirmed by means of direct numerical
simulations of the original non-autonmous DNLS model.

It is relevant to point out that excitations of the form considered
herein should in principle be accessible to state-of-the-art current
experiments in the optical (waveguide array) or atomic (Bose-Einstein
condensate) realm.
For instance, in the BEC context, 
such states could be obtained by considering a  2D  
array of $^{85}$Rb or $^7$Li created with  a deep two-dimensional optical lattice. In this case, the corresponding prototypical
mean-field model for the bosonic wavefunction $\psi$ would read
\begin{equation}
i\hbar\psi_t = -\frac{\hbar^2}{2m}\nabla^2\psi + g_{2D}|\psi|^2\psi + V_0(\sin^2(kx)+\sin^2(ky))\psi,
\end{equation}
where $g_{2D}$ is the effective nonlinear prefactor; see, e.g.,~\cite{emergent}.
$V_0$ characterizes the strength and $k$ the period of the optical lattice.
In the (superfluid) limit of a deep optical lattice, e.g. $V_0 \gg E_R=\hbar^2 k^2/2m$,  this equation reduces to  the DNLS model~\cite{ABKS}. For  $^{7}$Li one could use the Feshbach resonance technique at the external magnetic field of $B_0=738$G with the width $\Delta B =-170G $ \cite{Hulet} to modulate the 
interactions. The background scattering length can be taken as $a_{bg} \approx -20 a_B$, where $a_B$ denotes the Bohr radius. The time dependence of the scattering length $a_s (t)=a_{s0}+a_{s1}\cos(\omega t)$ follows from the variation of the  magnetic field in time according to
$$
a_s (B)= a_{bg}\left(1 - \frac{\Delta  }{B-B_0}\right).
$$
For  $B(t) =( 970 \pm 155 \cos(\omega t))$G, we have  $a_{s0}=-5a_B, a_{s1}= 10a_B$. Using suitable lattice parameters such as $2\pi/k = 0.5\mu$m
and $V_0 =15E_R$, we can ensure being in the regime of applicability of
the DNLS, enabling, under the above conditions, the experimental 
realizability of the higher-dimensional discrete compactons.


On the other hand, there are numerous interesting themes for future 
theoretical investigations. It would be interesting to examine
in more detail the origin of features that are fundamentally different
between the averaged model considered herein and the standard
DNLS model. These include among others (a) the restabilization of
in-phase configurations; (b) the absence of instabilities for
the discrete vortex configuration. Additionally, it would be especially
interesting to explore the approach of the model to the continuum 
limit of large coupling. It is well-known that collapse features
emerge as the DNLS model approaches the continuum limit~\cite{pgk_book},
although this may happen in unconventional ways for non-standard
discretizations of the model; see e.g.~\cite{herring}. It would
be intriguing to explore the properties of the present discretization
(and of the original non-autonomous model) as this limit is approached.
Finally, it would also be of interest and relevance to explore 
three-dimensional configurations in analogy to ones of the
standard DNLS~\cite{pgk_book} and to examine their stability properties.
Such studies  are currently in progress and will be reported in 
future publications.

\acknowledgements
M. S. acknowledges partial support from the Ministero dell'Istruzione, dell'Universit\'a e della Ricerca (MIUR) through a Programmi di Ricerca Scientifica di Rilevante Interesse Nazionale (PRIN) 2010-2011 initiative.
P.G.K.~gratefully acknowledges the support of
NSF-DMS-1312856, BSF-2010239, as well as from
the US-AFOSR under grant FA9550-12-1-0332,
and the ERC under FP7, Marie Curie Actions, People,
International Research Staff Exchange Scheme (IRSES-605096).
P.G.K.'s work at Los Alamos was partially supported by the
US Department of Energy. 
F.A. acknowledges the support from a senior visitor fellowship from CNPq (Brasil).

\end{document}